 \documentclass[preprint,review,3p,twocolumn]{elsarticle}

\usepackage{amssymb}
\usepackage{graphicx}
\usepackage{caption}
\usepackage{subcaption}
\usepackage{amsfonts}
\usepackage{amsmath}
\usepackage{balance}
\usepackage{amsthm}

\usepackage{comment}
\usepackage{enumerate}
\usepackage{morefloats}
\usepackage{balance}
\usepackage{hyperref}
\usepackage{xcolor}
\usepackage[modulo]{lineno}

 \linenumbers

\journal{Computer Networks}

\begin{document}

\begin{frontmatter}



\author[label1]{Constantinos Marios Angelopoulos}
\author[label2,label3]{Gabriel Filios}
\author[label2,label3]{Sotiris Nikoletseas}
\author[label4]{Theofanis P. Raptis\corref{cor1}}

\address[label1]{Bournemouth University, Poole, UK}
\address[label2]{Department of Computer Engineering and Informatics, University of Patras, Greece}
\address[label3]{Computer Technology Institute and Press ``Diophantus'' (CTI), Greece}
\address[label4]{Institute of Informatics and Telematics, National Research Council, Pisa, Italy}
\cortext[cor1]{Corresponding author: T.~P.~Raptis, \href{mailto:t.raptis@iit.cnr.it}{theofanis.raptis@iit.cnr.it}}


\title{Keeping Data at the Edge of Smart Irrigation Networks: \\A Case Study in Strawberry Greenhouses}


\author{}

\address{}

\begin{abstract}
Strawberries are widely appreciated for their characteristic aroma, bright red color, juicy texture, and sweetness. They are, however, among the most sensitive fruits when it comes to the quality of the end product. The recent commercial trends show a rising number of farmers who directly sell their products in the market and are more interested in using smart solutions for a continuous control of the factors that affect the quality of the final product. Cloud-based approaches for smart irrigation have been widely used in the recent years. However, the network traffic, security and regulatory challenges, which come hand in hand with sharing the crop data with third parties outside the edge of the network, lead strawberry farmers and data owners to rely on global clouds and potentially lose control over their data, which are usually transferred to third party data centers. In this paper, we follow a three-step methodological approach in order to design, implement and validate a solution for smart strawberry irrigation in greenhouses, while keeping the corresponding data at the edge of the network: (i) we develop a small-scale smart irrigation prototype solution with off-the-shelf hardware and software equipment, which we test and evaluate on different kinds of plants in order to gain useful insights for larger scale deployments, (ii) we introduce a reference network architecture, specifically targeting smart irrigation and edge data distribution for strawberry greenhouses, and (iii) adopting the proposed reference architecture, we implement a full-scale system in an actual strawberry greenhouse environment in Greece, and we compare its performance against that of conventional strawberries irrigation. We show that our design significantly outperforms the conventional approach, both in terms of soil moisture variation and in terms of water consumption, and conclude by critically appraising the costs and benefits of our approach in the agricultural industry.
\end{abstract}




\end{frontmatter}

\section{Introduction} \label{sec::intro}

Irrigation is crucial for agricultural production in order to ensure that greenhouse farmers can meet crop water demands. However, poor irrigation scheduling and inefficient utilization of water resources are two of several ubiquitous parameters restricting production in many agricultural regions. farmers can use sensed information such as light, moisture and temperature levels to modify irrigation schedules and avoid the risk of damaging crops \cite{7879140}. For example, soil sensors can be used to collect information on how water flows through the land and can be used to track changes in soil moisture, temperature, and levels of nitrogen and carbon. This allows the monitoring, optimization, and precise control of sensitive crops like strawberries and facilitates farmers in maximizing their crop production while maintaining very high quality in their end product.

A common approach for collecting large volume of agricultural data is based on the assumption that some network infrastructure is already present and is able to support the collection and delivery of all these data. Usually, data are pushed towards the cloud, which is intended to be the back-end aiming at processing and getting value from such data, as well as at controlling the field actuation devices. Moving all computing tasks to the cloud has been an efficient way to process data because there is more computing power in the cloud than in the devices at the network edge, where the field devices are deployed. Typically, for Internet of Things (IoT) enabled systems, this backbone is a wideband cellular network such as LTE. In the case of Smart Irrigation Network (SIN) environments an alternative may also include more localized wideband infrastructures, such as WiFi. In any case, an approach relying exclusively on global cloud providers to provide smart irrigation services has limitations mainly from two standpoints. On the one hand, wideband wireless networks may not provide sufficient bandwidth, particularly in rural areas, to support the data traffic demand. On the other hand, relying only on global clouds deployed may make SIN owners and operators lose control over their data, as those data are transferred to data centers without any control on behalf of the data owner. In addition, meeting the farmer's requirements in terms of storage and computation capacity may have a significant impact on the cost incurred to the farmer for ICT services, which, if reduced, could be more profitably invested in the core production process. Consequently, keeping the data at the network edge yields shorter response times, more efficient processing and actuation, less pressure on the backhaul network and more robust data ownership guarantees \cite{7469991}.

The concerns regarding moving the agricultural data away from the edge of the network are discussed in depth at the position paper ``Cloud of Things in Smart Agriculture: Intelligent Irrigation Monitoring by Thermal Imaging'', authored by M.~Roopaei et al., \cite{7879140}. The authors examine key technical and legal issues and requirements supporting the use of Cloud of Things for managing water source-related data. They present the advantages of cloud-based approaches for smart irrigation; however, they identify the security and regulatory challenges which come hand in hand with sharing the crop data with third parties, outside the edge of the network. Regarding the security aspects, the authors state that, unfortunately, cloud-based smart irrigation systems potentially have more attack vectors (e.g., hardware,  firmware, and applications running on Cloud of Things devices) that can be remotely exploited by attackers, particularly during early stages and in comparison to traditional, isolated, irrigation systems (which keep the data at the edge). Regarding the regulatory aspects (e.g., data protection and the Internet governance), the authors state that the time required to develop an appropriate legal and regulatory framework is significantly longer than the time it takes to develop the next-generation SIN cloud systems. 

Strawberries are widely appreciated for their characteristic aroma, bright red color, juicy texture, and sweetness. The recent commercial trends show a rising number of farmers which directly sell their products in the market and are more interested in using smart solutions (like SINs) for a continuous control of the quality related factors\footnote{FA.MO.S.A. sarl: \url{http://www.famosasrl.com/en/}}. In fact, strawberries cultivated in greenhouses are very susceptible to water irrigation amounts which are very important both during the first months after planting and before harvesting. Water amounts have to be constantly maintained within optimum ranges, in order to avoid loss of product which otherwise can reach up to the 80\% of the yield, caused by the presence of misshapen, plant collapsed and small fruit. Farmers need to know the level of greenhouse temperature and soil water content many times a day, in order to make decisions about temperature management and water supply\footnote{Innovation Centre for Sensor and Imaging Systems (Censis): \url{https://censis.org.uk}}.

\subsection{Contributions and roadmap}

In this paper, contrary to the traditional cloud-based solutions, we propose a decentralized smart irrigation approach for strawberry greenhouses, the core idea of which is to keep the agricultural data within the range of the edge of the network. Decentralized data management, a key component of edge computing, can be a very suitable approach to cope with the aforementioned challenges. In the context of SINs, one could leverage the set of nodes present at the edge of the network to distribute functions that are currently being implemented in remote data centers \cite{s18082611}. More specifically, we adopt a three-step methodological approach which leads to our three following contributions:


\begin{itemize}
\item In the first step, we develop a small-scale SIN prototype with off-the-shelf hardware and software equipment. We test the performance of this prototype on different kinds of plants in order to investigate and evaluate the feasibility of our ideas and to gain some insights for larger scale deployments. We focus on maintaining small variability in terms of soil moisture levels compared to the conventional timed watering approach.
\item In the second step, using the experience gained from the small-scale prototype, we introduce a SIN reference architecture, specifically targeting edge data distribution for strawberry greenhouse applications. We analyze the different electromechanical and networking components needed for implementing such a SIN and we tailor the architecture to the modern strawberry greenhouse requirements.
\item In the third step we implement a full-scale SIN system in an actual strawberry greenhouse environment in Greece, by adopting the proposed reference architecture. Building upon the lessons learnt during the small-scale pilot of the first step, and the evident need for a hardware design that would facilitate interfacing different types of sensors and actuators with IoT development platforms, while at the same time providing a sufficient degree of physical robustness, we propose a new hardware design for sensing and actuation in SINs, the control cube. We implement the hardware design and we setup the system in two greenhouses:  In the first greenhouse the process is supervised by the farmer. The second greenhouse is managed by the developed SIN system. Based on the final results, our SIN significantly outperforms the conventional approach both in terms of soil moisture variation and in terms of water consumption.
\end{itemize}

The roadmap of the paper is the following: In section \ref{sec::ref}, we present some representative works related to this paper. In section \ref{sec::small}, we design and evaluate the performance or our small-scale SIN prototype. In section \ref{sec::arch}, we introduce the reference architecture which can serve as the foundation of SIN systems targeting strawberry greenhouses. In section \ref{sec::big}, we present the implementation and the performance evaluation of the full-scale SIN system in real strawberry greenhouses environment by adopting the proposed reference architecture. Finally, in \ref{sec::conc} we conclude the paper. We also critically appraise the costs and benefits of our approach in the agricultural industry.

\section{Related Works} \label{sec::ref}

In this section, we present some selected previous research works. At first, we present some recent works on cloud-based irrigation solutions (which prefer to push the data in the cloud or to third party systems for conducting smart irrigation). Then, we present some interesting past works focusing on the exact topic of strawberry smart irrigation. Finally, we conclude the section with some IoT device and system designs targeting smart agriculture in general.

\emph{Cloud-based solutions}. Researchers at Colorado State University have created an online evapotranspiration-based irrigation scheduling tool called Water Irrigation Scheduling for Efficient Application (WISE) that uses the soil water balance method and data queries from Colorado Agricultural Meteorological Network (CoAgMet) and Northern Colorado Water Conservation District (NCWCD) weather stations \cite{BARTLETT2015127}. To expedite and mobilize required user interaction with the software interface, a smartphone app has been developed that allows users to quickly view their soil moisture deficit, weather measurements, and the ability to input applied irrigation amounts into WISE. In \cite{7389138}, the authors propose and evaluate on a real deployment a cloud-based Wireless Sensor Network (WSN) communication system. This solution monitors and controls a set of sensors and actuators, respectively, to assess plants water needs. A remote web service is employed to optimize the system with weather knowledge.

\emph{Strawberry smart irrigation}. In \cite{LOZANO201644} the authors measure crop evapotranspiration in Do\~{n}ana National Park, Spain, of two commonly used strawberry cultivars using drainage lysimeters, estimate their crop coefficients, and evaluate their irrigation efficiencies, crop yields and water productivities. Then, they develop a simple Android application which facilitates the manual irrigation scheduling in the strawberry sector, highlighting that efficient water use is in fact possible in the strawberry production. In \cite{8229715}, the authors analyze the agro-climatic variables and the water requirements by evapotranspiration for strawberry plantations in the commune of San Pedro, Chile. They use correlation analysis, principal components analysis, $k$-means of the time series of the agro-climatic variables and evapotranspiration as methodological enablers. They classify the periods with need of water to implement a water balance controller.

\emph{IoT designs for smart agriculture}. In \cite{8058860}, the authors discuss how critical is the moisture level for agriculture as is directly related to the plants water stress. Among others, measuring the temperature difference between the leaf and the air is a way to detect it. This paper introduces a tag that consists of a wireless leaf temperature sensor, a low power microcontroller, a sensor board and a module for wireless communication. This tag communicates using backscatter radio principles on ambient FM station signals using AM modulation. Similarly, in \cite{8423620} a low-cost, and low-power system for leaf sensing using a new plant backscatter sensor node/tag is presented. As previously, a tag measures the temperature difference between the leaf and the air and then, communicates remotely with a low-cost software-defined radio reader. The node consists of the sensor board, a microcontroller, an external timer, and an RF front-end for communication while all of them are supplied by a solar panel without the need of a battery. The work conducted in \cite{4382216} presents the results of a real deployment that is based on WSN and is responsible to monitor and control a number of environmental attributes. The system is deployed in greenhouses with melon and cabbage and is used to monitor the growing process of them as well as manage the environmental conditions of the greenhouses. In \cite{4340413} the authors present software and hardware design solutions based on Zigbee communication to accommodate monitoring and control capabilities for agriculture applications, while in \cite{inproceedings} information concerning the Zigbee specifications are provided, as well as its benefits in agricultural use. Irrigation plays a critical role in agriculture, as it consist a quality and resource cost factor. In papers \cite{4457920,7133593,7133592,7930391} the authors deal with the efficient use of water as well as its nutrient-rich levels. WSNs combined to both decision support system and prediction models aim to support smart management and schedule optimization. Finally, \cite{8612441,8644215,7219582} correspond to smart farming with the IoT concept. Particularly, cheap hardware is used to built an intelligent system which improves productivity, experience and management towards modern farming time.

Different from those approaches, we envision SINs in which the farmer and the data owners are able to fully control their data, as well as process and analyze them within the range of the edge of the network. There have been some recent works in the general IoT-enabled agriculture literature which try to address related considerations at the edge of the network. Those works are focusing on different application areas and target objectives. For example, in \cite{8644296}, the authors propose an edge master-slave machine learning architecture, in which pre-trained lightweight machine learning models at the edge identify the origin of the incoming packets based on the long-term learned collective variations of the sensorial values from the slave node. However, this architecture focuses on device energy efficiency, as the objective is to increase the amount of sensor data transmitted between the slave to master nodes with significant energy savings over longer duration. Another example is \cite{BU2019500}, in which the authors also adopt edge-based operations, but, contrary to our approach, combine them with a hybrid cloud-based scheme. Specifically, they present a smart agriculture IoT system based on deep reinforcement learning which includes four layers, namely agricultural data collection layer, edge computing layer, agricultural data transmission layer, and cloud computing layer. The presented system integrates some advanced information techniques, especially artificial intelligence and cloud computing, with agricultural production to increase food production. Finally, in \cite{Chi2019}, the authors employ an edge agrarian deployment of sensor nodes with RFID, WiFi and Bluetooth modules. However, the goal of this paper is not to suggest an architectural approach like our paper, but to make a very thorough quantitative analysis on the theoretical maximum collision time and collision probability of WiFi or Bluetooth network with RFID interferers.

\section{Small-scale prototype implementation and initial results} \label{sec::small}

The first step in our methodological approach has been to design and implement a small-scale prototype implementation\footnote{A preliminary version of the concepts and results reported in this section has been presented in \cite{Angelopoulos2011}.}. The purpose has been dual: (a) for the prototype to act as a proof-of-concept system demonstrating the potential gains of a full-scale system in terms of efficient irrigation and water savings as well as its agility in dynamically adapting the irrigation process to the actual agricultural needs and (b) to gain early insights, before the actual full-scale deployment, on the design process of a SIN which keeps the data at the edge of the network. 

Our pilot took place in south-western Greece during early summer time, when the climate is characterized by fairly dry weather conditions with temperatures averaging between low and high 30$^{o}$C. We chose this place and time of the year due to the close environmental conditions they bear to the ones observed at the strawberry greenhouses in which we later deployed the full-scale system implementation (section \ref{sec::big}). The developed system consisted of wireless sensor motes, soil moisture sensors, mote-driven electro-valves that controlled water flow towards the plants, and a Java application running on a PC collecting data from the sensor network and storing them in a MySQL database. For the purposes of this pilot, and due to the fact that we did not yet have access to large-scale strawberry plantations, three plant pots where used, each one containing a different plant with highly diverse watering needs: (a) geranium, which has very limited watering needs (once per week), (b) lavender, which under normal weather conditions, has moderate watering needs (three times per week), and (c) mint, which requires regular watering (in high temperatures during summertime, even twice per day). 

The soil moisture of each pot was monitored by a sensor mote equipped with soil moisture sensor. Similarly, irrigation for each pot was controlled by a corresponding mote-driven electro-valve independently from the other pots. When soil moisture levels of a pot were dropping below a predefined threshold, the mote which was monitoring its soil moisture was informing the mote driving the corresponding electro-valve to start watering the pot. Once the soil moisture returned to normal levels, the soil monitoring mote would signal the electro-valve mote to stop watering the pot. During the operation of the system, all measurements and actions taken by the motes were forwarded by the corresponding motes to the mote acting as a gateway for the rest of the motes by being connected to a PC over the serial port (circa 25 meters away). Received sensory measurements and actions taken (i.e., motes driving the watering valves) were logged in a MySQL database for further processing. Note that all data, as well as all actuation decisions taken were localized inside the edge of the deployment.

\subsection{Pilot setup}

\begin{figure*}[t!]
\begin{subfigure}{.49\textwidth}  
\centering
\includegraphics*[scale=0.39]{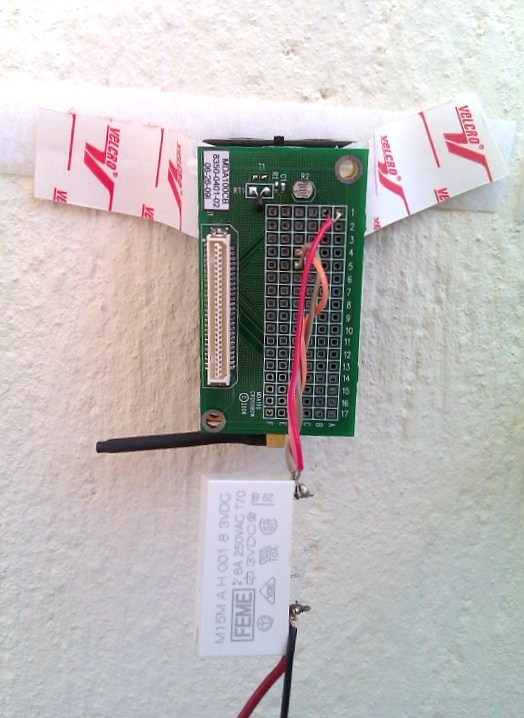}
\caption{Relay shouldered on the MDA100CB board of the IRIS mote.}
\label{fig:iris-relay}
\end{subfigure}
\begin{subfigure}{.49\textwidth}  
\centering
\includegraphics*[width=\textwidth]{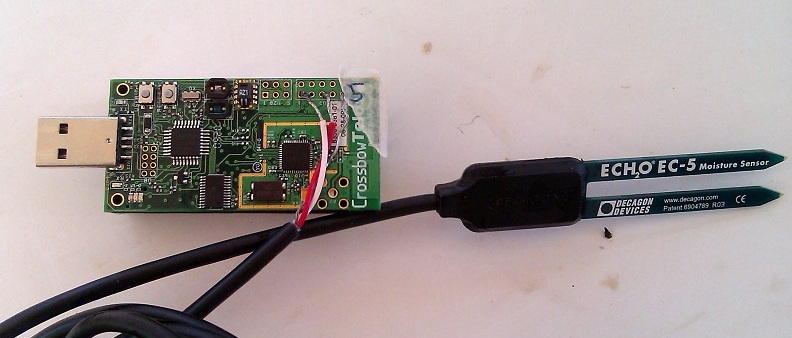}
\caption{The EC-5 soil sensor shouldered on a TelosB mote.}
\label{fig:telos-soil}
\end{subfigure}

\begin{subfigure}{.49\textwidth}  
\centering
\includegraphics*[scale=0.29]{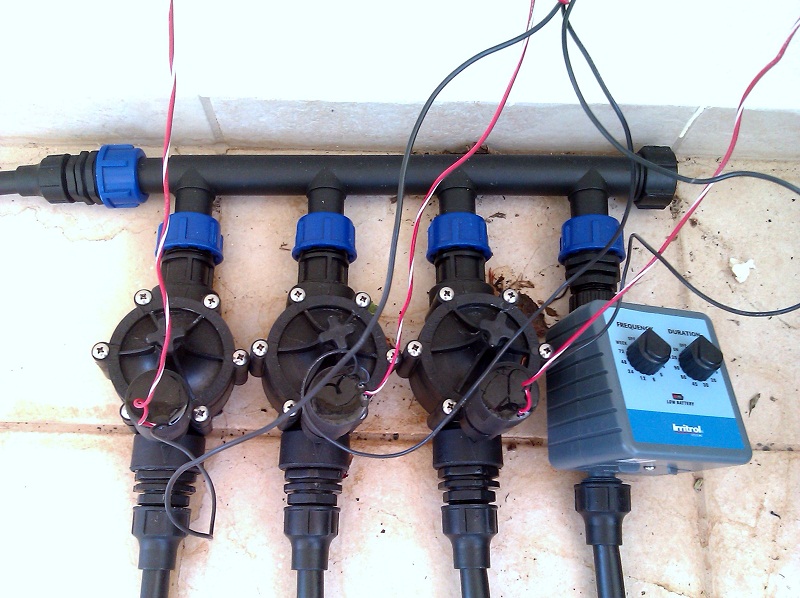}
\caption{Installed electrovalves and common irrigation programmer. The red and black cables of the solenoids are connected in series with the relay and the external power source (batteries or AC-DC converter).}
\label{fig:valves-common}
\end{subfigure}
\begin{subfigure}{.49\textwidth}  
\centering
\includegraphics*[scale=0.29]{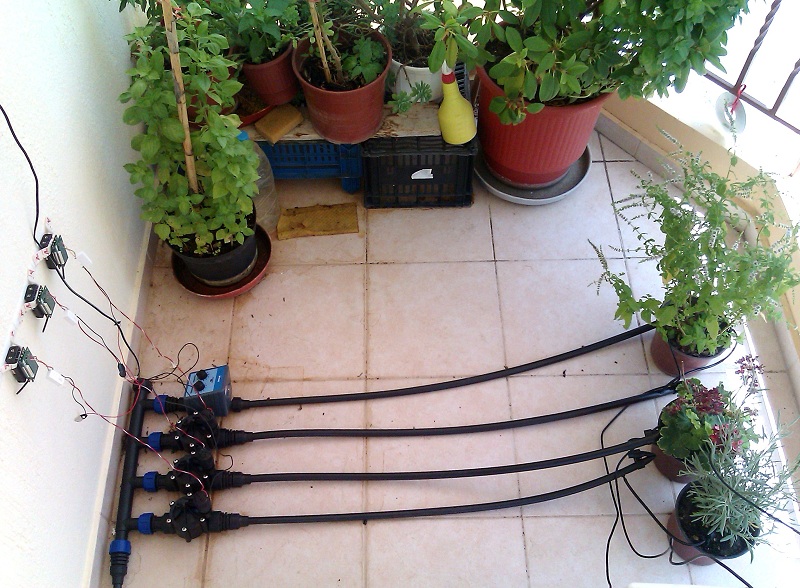}
\caption{The entire small-scale SIN.}
\label{fig:garden}
\end{subfigure}
\caption{Sensing and actuating devices used at the installation.}
\end{figure*}

For our implementation we used two mote platforms; the TelosB and the IRIS motes. Both of them are IEEE 802.15.4 compliant, small, light weight and when using energy saving protocols can be powered with two AA batteries for several months, even years. These characteristics make them ideal for our pilot as they can easily be deployed everywhere while being independent of power installations. Furthermore, the IRIS platform was combined with the MDA100CB sensor and data acquisition board which has a precision thermistor, a light sensor/photocell and a general prototyping area. This prototyping area was used in order to connect a relay, through which the IRIS motes were able to control the electrovalves (Fig.~\ref{fig:iris-relay}). Finally, it is worth mentioning that, due to their highly constrained nature, the TelosB and the IRIS motes have been ideal in showcasing that an efficient IoT-enabled SIN can be developed by using low-spec, cost-efficient IoT devices that are commercially available.

The EC-5 soil moisture sensor by Decagon was used for soil monitoring (Fig.~\ref{fig:telos-soil}). It consists of a cable, which on one end has two prongs and on the other end has 3 wires. The prongs are pushed inside the potting soil and the three wires of the other end are connected to the 10-pin expansion connector of TelosB motes. The bare wire is connected to the ground pin, the red one is connected to the ADC channel pin (programmed as input) and the white wire is connected to the VCC pin. This sensor had to be used along with TelosB motes as it provides 12-bit data, while the IRIS mote has a 10-bit ADC. The communication between the TelosB mote and the soil moisture sensors was established by the corresponding components and interfaces provided by the TinyOS operating system which was used to flash the motes.
 
In order to control the irrigation process we used solenoid valves provided by Irritol (Fig.~\ref{fig:valves-common}). A solenoid valve is an electromechanical valve that is controlled by an electric current. The electric current runs through a solenoid, which is a wired coil wrapped around a metallic core. The solenoid creates a controlled magnetic field when electrical current is passed through it. This magnetic field affects the state of the solenoid valve, causing the valve to open or close. The electric valves operate on a 9V battery.

In order for the IRIS motes to be able to drive the valves we used relays that take as input a 3V DC current and can control circuits of up to 250V AC. Due to the low current IRIS motes can provide as output, the relay is driven by using two digital outputs of the prototyping area of the MDA100CB board. Then, the relay is connected in series with the electro-valve and an external 9V battery. This way, when the IRIS mote triggers the relay closing the circuit, the electro-valve is activated and the irrigation process begins. 

In order to have a benchmark for the performance evaluation of our deployed system, we also installed a common irrigation programmer representing the conventional irrigation methods (Fig.~\ref{fig:valves-common}). This programmer can be set to irrigate at regular time intervals for fixed periods of time. In our case, we set it every second day for half an hour. The entire SIN is displayed at Fig.~\ref{fig:garden}.


The firmware for the IoT motes has been developed with the TinyOS operating system. TinyOS provides a component based architecture and forms an event-driven operating system. The motes where assigned unique mote IDs and were programmed so that for each pot each mote forwards sensory data on soil moisture to the IRIS mote controlling the corresponding watering electro-valve. 

\begin{figure*}[t!]
\begin{subfigure}{.49\textwidth}  
\centering
\includegraphics*[width=\columnwidth]{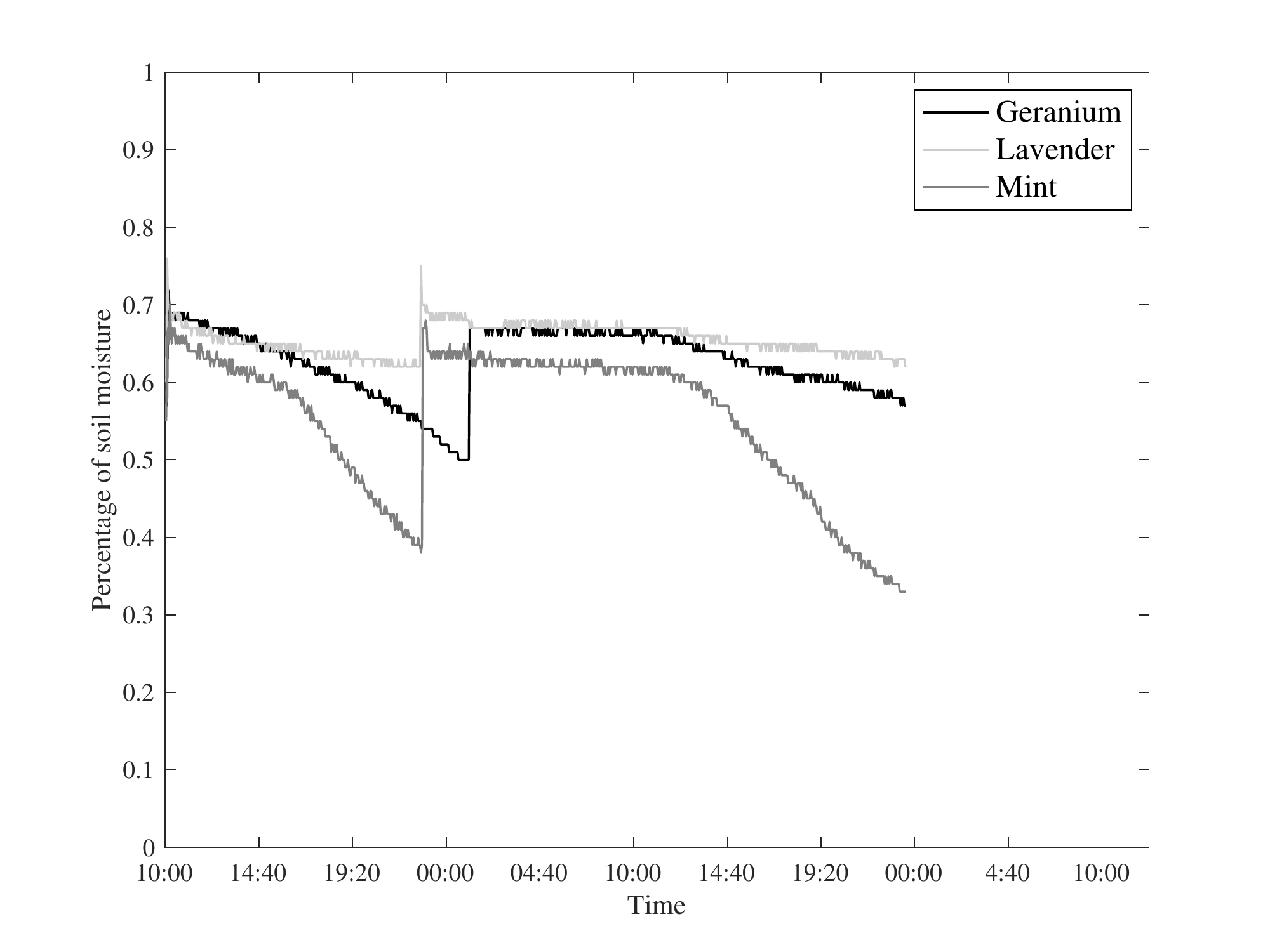}
\caption{Common irrigation.}
\label{fig:manual}
\end{subfigure}
\begin{subfigure}{.49\textwidth}  
\centering
\includegraphics*[width=\columnwidth]{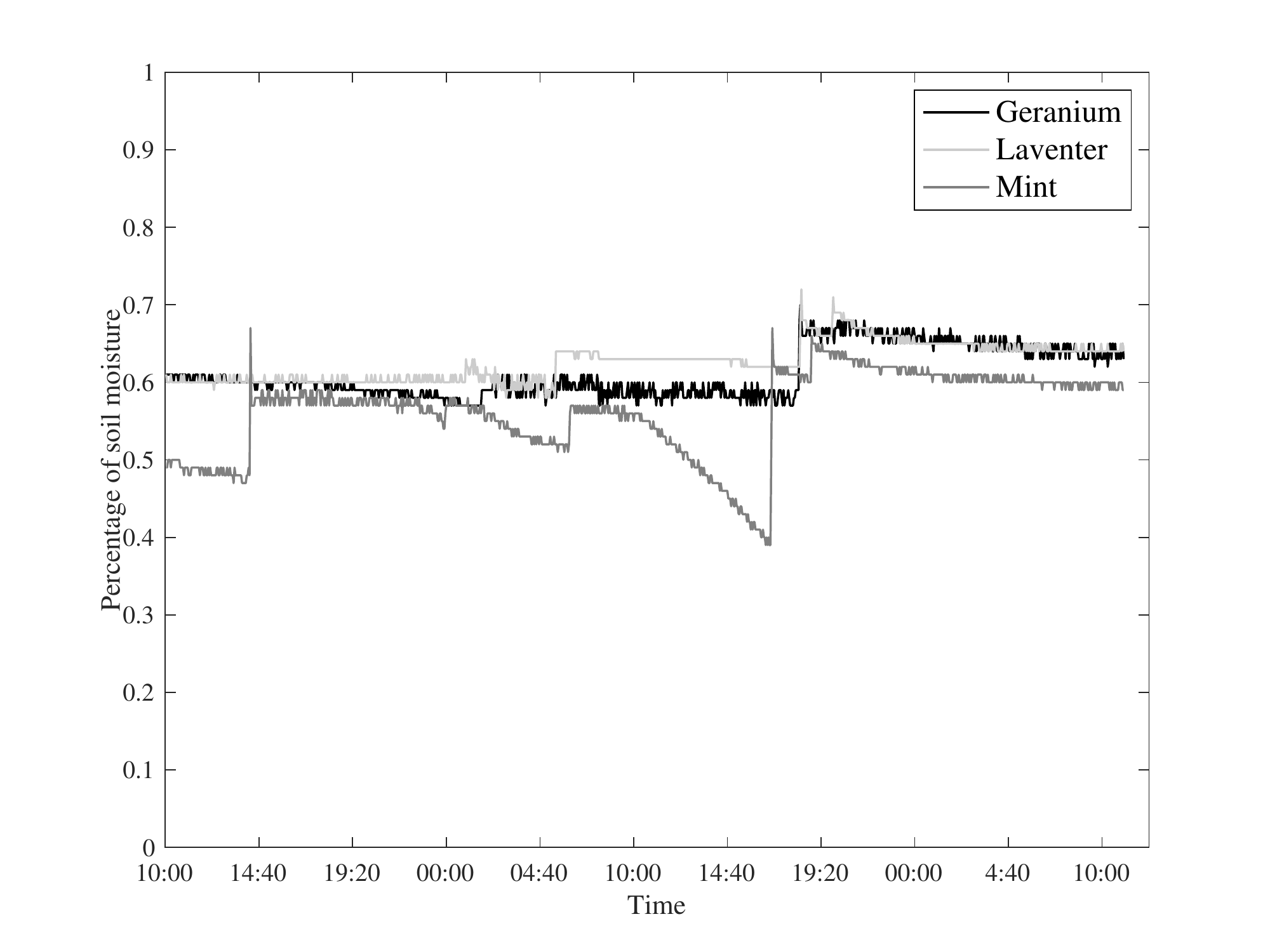}
\caption{Smart irrigation.}
\label{fig:smart}
\end{subfigure}
\caption{Soil moisture over time.}
\end{figure*}

The pilot makes use of three TinyOS applications. The \textit{SoilSensorApp} (runs on TelosB motes) uses the \textit{humidity\_msg} and the \textit{valve\_msg} message types defined in the \textit{SoilSensor.h} header. The first type is used in order to send moisture values to the Sink to be logged at the MySQL database. The second type is used in the TelsoB - IRIS communication. If the value of soil moisture gets below a lower bound, defined by variable \textit{low\_lim}, or above an upper bound, defined by variable \textit{upper\_lim}, then a \textit{valve\_msg} message with an appropriate value is sent to the IRIS mote controlling the corresponding electro-valve to set its state to be on or off. 

The second TinyOS application is the \textit{ValveApp} (runs on IRIS motes). With this application running, if the mote receives a \textit{valve\_msg} message, it reads its payload and accordingly sets or clears the output ADC pins. This way it can trigger the relay connected to these pins on/off; thus driving the electro-valve.

\subsection{Performance evaluation and insights}

We evaluated the performance of the traditional irrigation scheme, using a common irrigation programmer, and that of the smart irrigation scheme, using our deployed SIN. We let each system to water for two days the three pots, each one containing a different plant. The plants where chosen in order to have diverse water needs. Sorted in descending water needs, the plants were a geranium, a lavender and a mint. The season the experiments were conducted was early summertime and the temperature was around 36$^{o}$C during the day and 30$^{o}$C during the night. 
Figs. \ref{fig:manual} and \ref{fig:smart} show the soil moisture of each pot over time for the two irrigation schemes respectively. We note that the pot containing the mint, that has the highest water needs from all three plants we used, depletes moisture form its containing soil much faster. On the contrary, the geranium absorbs moisture much slower, therefore the soil dries out at a lower rate.

When using a common irrigation programmer there exist great variations in the concentration of soil moisture. The soil tends to dry out and then is flooded with water causing an almost vertical increase of the soil moisture values. These extreme variations are not to the benefit of the plants, as they require soil moisture to remain at a given level. Furthermore, dried out soil causes great amount of water to pour away as  it cannot withhold water to the same degree as even lightly moist soil.

\begin{figure*}[t!]
\centering
\includegraphics*[width=\textwidth]{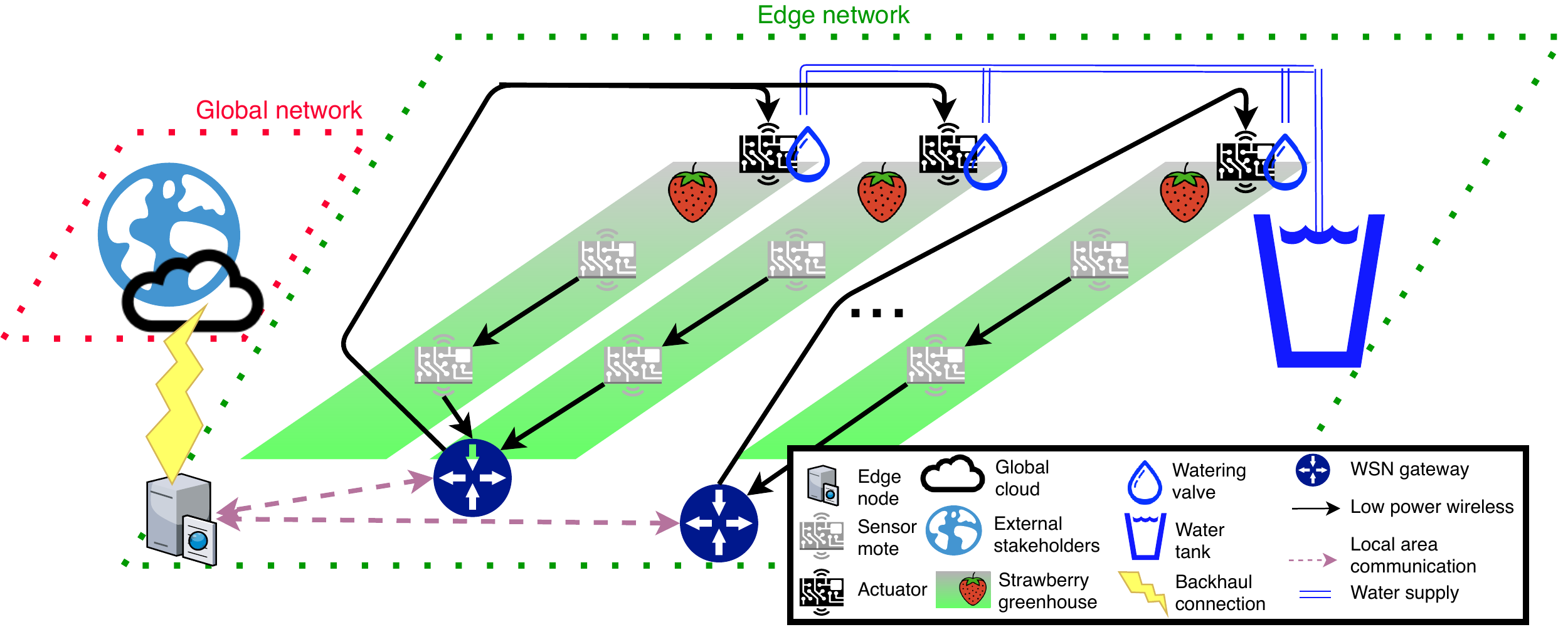}
\caption{Proposed architectural design for SINs applied to strawberry greenhouses.}
\label{fig:arch}
\end{figure*}

On the other hand, our SIN manages to maintain soil moisture at the same level. It dissipates less water and it provides an irrigation scheme that is adaptable to the specific watering needs of each plant. The most important feature is the fact that by constantly monitoring the moisture levels, the system is able to dynamically adapt to the current environmental conditions. Whether there are fluctuations in temperature or sunlight variations, the system will adjust the irrigation process accordingly so as to maintain a constant level of soil moisture. Due to the small scale of the pilot and the short evaluation period, there were no safe conclusions regarding water consumption. However, the pilot has been successful in demonstrating that it is possible to achieve significant gains with cost-effective, commercially available components.  

\section{An architectural design for the strawberry greenhouses case} \label{sec::arch}

The second step in our methodological approach has been to design a reference architecture which can serve as the foundation of SIN systems targeting strawberry greenhouses. The proposed architectural design is displayed at Fig.~\ref{fig:arch}. The core assumption of our design lies on the fact that we opt for keeping the data within the range of the edge network (green-dotted rectangle in Fig.~\ref{fig:arch}) and not share them with the global network (red-dotted rectangle in Fig.~\ref{fig:arch}). Typically, strawberry plantations consist of parallel, long, rectangular greenhouses (green/gray-colored rectangles in Fig.~\ref{fig:arch}). Each greenhouse consists of multiple lines with strawberry plants, where each line has its own watering hose. The watering hoses of each greenhouse are connected to the main water supply through a valve. If a valve is open, the water supply is able to pass through the watering tubes to the greenhouse and to the watering hoses of the strawberries. In traditional deployments, the valve opening/closing schedules are configured either manually (by the farmer), or through common irrigation programmers which can be set to irrigate at regular time intervals for fixed periods of time. However, in our SIN architecture, we replace this conventional functionality with actuators which are able to both control the valves and communicate with other devices via low power wireless communication. In order to acquire the necessary readings regarding ambient conditions in the greenhouse, like soil moisture, the architecture provisions the installation of sensor motes in the greenhouses. Each greenhouse should have more than one sensor mote for more accurate estimations. The sensor motes can communicate with other motes via low power wireless communication, and can ultimately create multihop low power networks, by using standardized networking protocol stacks. The sensed data are then propagated to gateways positioned near the greenhouses. The number and the exact positioning of the gateways depends on the specific settings and requirements of each individual plantation. The gateways relay the data to a local edge node, via local area communication. The edge node is the main control unit of the installation and is able to analyze agricultural data, run valve scheduling algorithms, and perform also other tasks. One of the most important tasks is the real-time control of the greenhouse actuation. The edge node, while receiving real-time agricultural data originated from the sensor motes and relayed through the gateways, can communicate back to the gateways control data via local area communication, which in turn are relayed to the actuators via low power wireless communication. This way, the actuators are able, in real-time, to open and close the electromechanical valves and consequently schedule the irrigation process. Note that, as shown in Fig.~\ref{fig:arch}, there is the option for the edge node to be connected also to the global network if needed by the requirements. This backhaul Internet connection can be implemented through emerging 5G technological enablers and can provide additional functionalities, aligned to existing cloud-based solutions. However, as described in section \ref{sec::intro}, the transition from edge-based to cloud-based communication, comes with various additional restrictions and vulnerabilities \cite{8764545}: susceptibility to potential intrusion threats \cite{7879140}, higher end-to-end data access latency \cite{Lucas2018}, as well as dependence on third-party service providers \cite{Hosseini2019}.

\section{Full-scale system implementation} \label{sec::big}

The third step in our methodological approach has been to implement a full-scale SIN system in real strawberry greenhouses environment by adopting the proposed reference architecture. Building upon the lessons learnt during the small-scale pilot presented in section \ref{sec::small}, the need was evident for a hardware design that would facilitate interfacing different types of sensors and actuators with IoT development platforms, while at the same time providing a sufficient degree of physical robustness. Furthermore, for the full-scale system design to be scalable and energy efficient from a networking perspective is also of great importance; particularly when considering that in agricultural-related use cases any ICT infrastructure will most probably need to be battery operated. 

\subsection{Pilot setup} \label{sec::operation}

\begin{figure}[t!]
\centering
\includegraphics[trim={0 0 0 3.5cm},clip,angle =270,width=0.9\columnwidth]{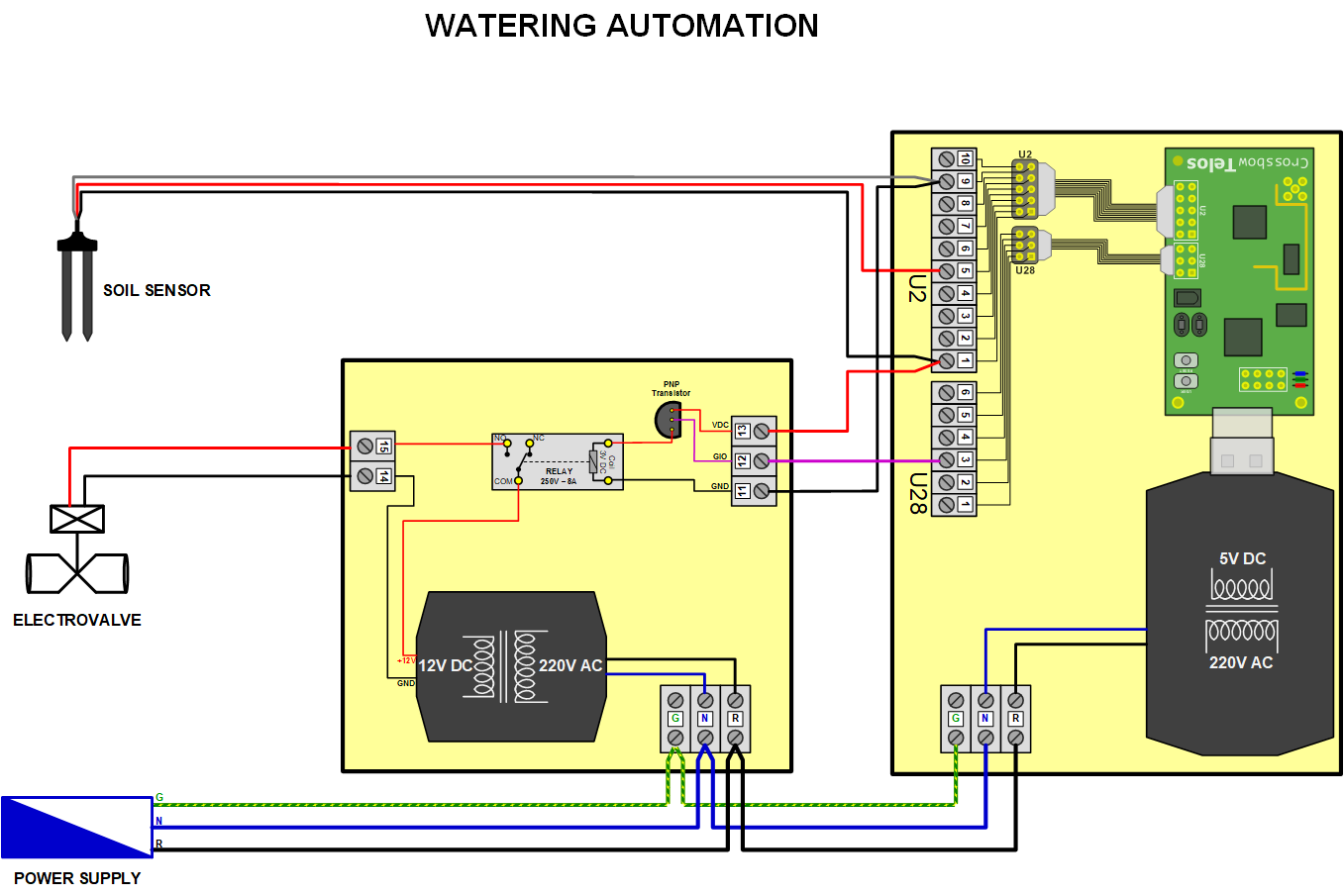}
\caption{Blueprint of our control cube (sensor) hardware implementation}
\label{fig:ccschetch}
\end{figure}

We chose to keep as our main hardware platform the TelosB sensor motes, as in section \ref{sec::small}, and further increment the hardware design \cite{6655569}. Although this platform is characterized by severe limitations in terms of computing power, communication range and memory, its most important aspect is that it is economic, and it can be easily modified and connected to other hardware components \cite{6846177}. In order to facilitate (and somehow standardize) the interfacing of the TelosB motes with the various sensors and actuators of the large-scale SIN, we have developed an additional hardware component, the \emph{control cube}, and adapted it to SIN applications. The control cube is an interconnection board consisting of AC/DC transformers, electronic relays, resistors, diodes and fuses. It is installed in series with the controlled device, and it is \emph{device and vendor independent}. We use the general I/O (GIO) pins of the TelosB platform to drive the electronic relays controlling the power supply circuit of the controlled device. Our design also offers the ability to choose between automated control via the sensor network or manual control via ordinary switches. The on-board sensors of the TelosB mote platform can be used to monitor ambient environmental conditions (i.e., temperature, relative air moisture and luminance levels). Similar to the small-scale pilot of section \ref{sec::small}, we used electromechanical valves provided by Irritol in order to control the irrigation process, and EC-5 soil moisture sensors by Decagon \cite{ec-5} in order to perform the soil moisture monitoring. The control cube design blueprint is displayed at Fig.~\ref{fig:ccschetch}. Note that the design of the control cube totally conforms to the relevant EU safety regulations. 

\begin{figure}[t!]    
     \centering
        \includegraphics[width=0.5\textwidth]{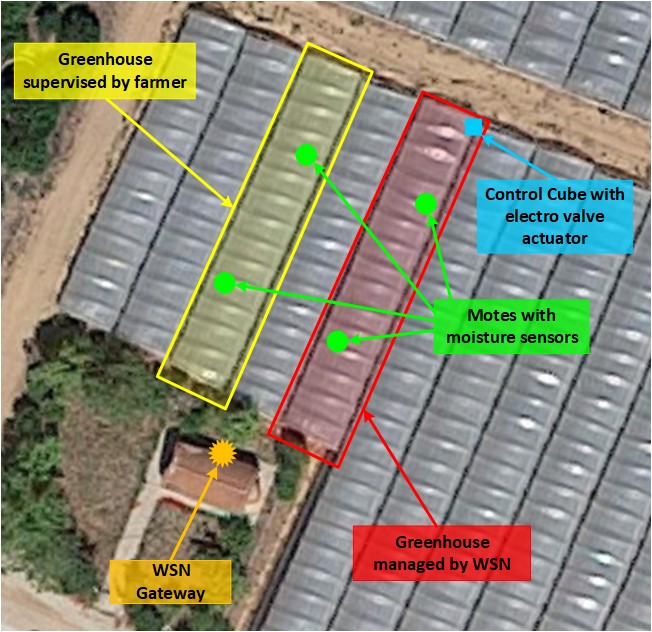}
        \caption{Topology of wireless sensor network deployed in the greenhouses.}
        \label{fig:facility}
\end{figure}


The SIN that was designed and developed, follows the architectural design presented in section \ref{sec::arch}, and was piloted during the strawberry harvest season (from May to July) in a strawberry growing facility in western Peloponnese, Greece which covers an area of 400$m^2$. The infrastructure's specifications among others include sandy soil cultivation in greenhouses at the root of the plant, as well as water of medium hardness and 6.5$pH$. The irrigation of the crop is done with water pumped by drilling, which is stored in a tank and then is fed back to the crop through a surface irrigation system. Both the selection of the crop and the trial period of the system, were made based on the sensitivity of the strawberry plant to the fluctuations of water supply.

\begin{figure*}[t!]
\begin{subfigure}{.47\textwidth}
     \centering
        \includegraphics[height=9cm,keepaspectratio]{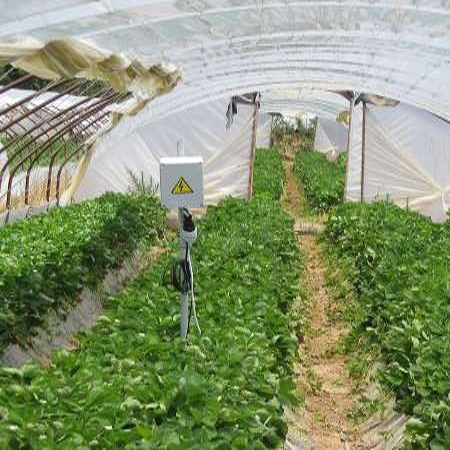}
        \caption{Control cube (sensor) measuring soil moisture.}
        \label{fig:1}
\end{subfigure} 
\begin{subfigure}{.47\textwidth}   
     \centering
        \includegraphics[height=9cm,keepaspectratio]{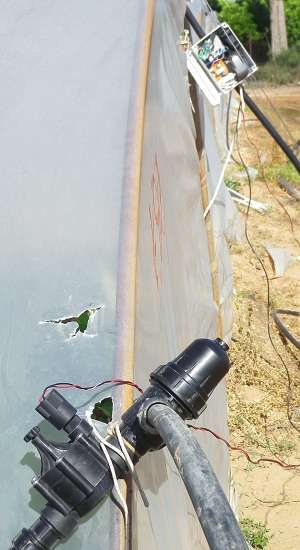}
        \caption{Control cube (actuator) connected to electrovalve.}
        \label{fig:2}
\end{subfigure}
\caption{Sensing and actuating devices used at the strawberry greenhouse.}
\label{fig:devices}
\end{figure*}

In particular, we used two greenhouses for comparison reasons, as shown in Fig.~\ref{fig:facility}. In the first, for benchmarking purposes, we used the traditional way of monitoring irrigation process supervised by the farmer, whereas in the second we adopted an irrigation process which was managed automatically through a deployed SIN following the architecture of section \ref{sec::arch}. The components that are used in the SIN consist of sensor motes in order to monitor the soil moisture levels and control cubes in order to actuate on the water supply in the greenhouses. 

Each greenhouse consists of 4 lines with strawberries plants, where each line has its own watering hose. The four watering hoses of each greenhouse are connected to the main water supply through a valve. Usually, in large plantations like this one, the farmers are leaving these valves open and organize the greenhouses in sectors with a central valve for each sector, through which they control the irrigation process of the whole sector.  Until now, the farmers were watering all the greenhouses with a schedule for the whole cultivation, based on their experience, ignoring special conditions of each greenhouse. That was usually resulting in a production variation for different greenhouses both on quality of the fruits and the production quantity.

\begin{figure*}[t!]
\begin{subfigure}{.47\textwidth}
  \centering
  \includegraphics[width=\textwidth]{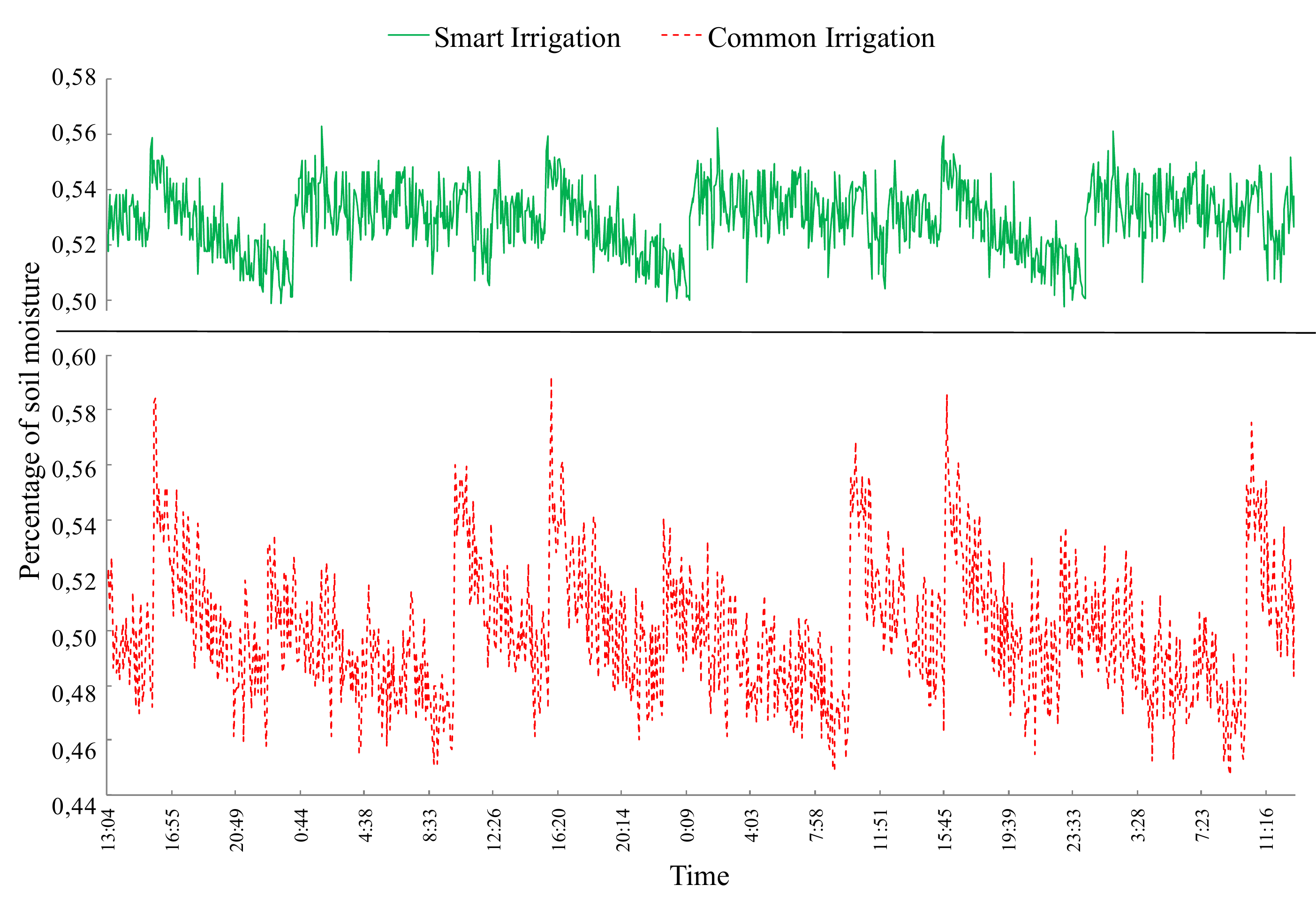}  
  \caption{Soil moisture.}
  \label{fig:moisture}
\end{subfigure}
\begin{subfigure}{.47\textwidth}
  \centering
  \includegraphics[width=\textwidth]{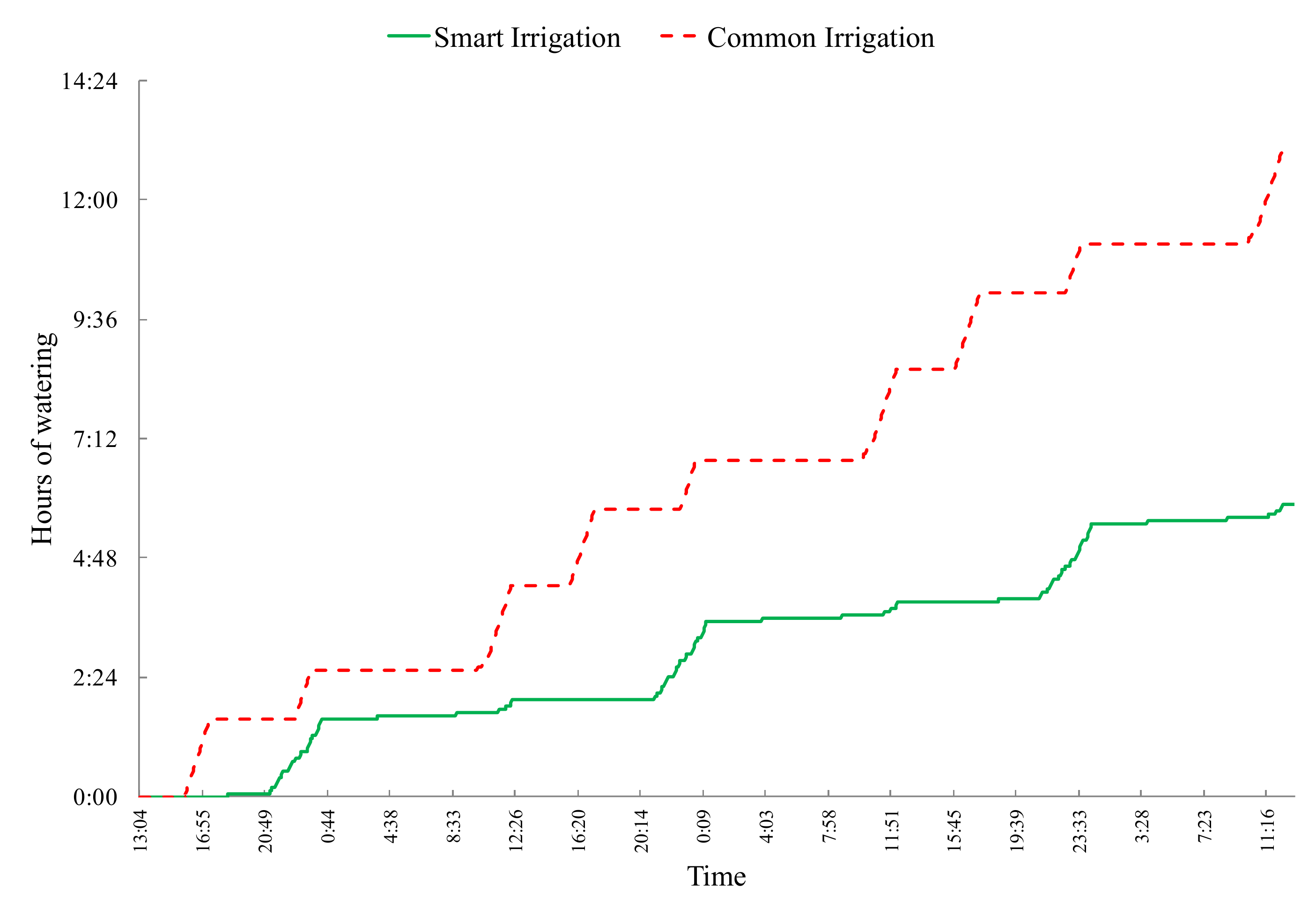}  
  \caption{Water consumption.}
  \label{fig:water}
\end{subfigure}
\caption{Soil moisture and water consumption over time for a period of 3 days.}
\label{fig:measurements}
\end{figure*}

We designed the exact deployment of the SIN with the corresponding sensors and actuators, for monitoring the conditions into the greenhouse and automate watering when it is needed, as shown in Fig.~\ref{fig:facility}. First, we decided to deploy sensing control cubes in two neighboring greenhouses to monitor the conditions inside these greenhouses and measure the soil moisture at the roots of the plants (Fig.~\ref{fig:1}). The soil moisture levels are important in order to decide and answer the questions ``when'' and ``how much'' water is needed. In each greenhouse, we deployed two sensor motes in different lines to monitor the local conditions (Fig.~\ref{fig:2}). In the next step, we replaced the manual valve in the one greenhouse with an electrovalve that could be controlled by the actuating control cube. The control cube, as an actuator described by the reference architecture of section \ref{sec::arch}, communicates wirelessly with the gateway of the network and is able to open or close the water supply of the specific greenhouse. A single gateway was installed as shown in Fig.~\ref{fig:facility}, and is used to perform the communication between the sensor motes/control cube and the local edge node located nearby, at the plantation control center. This gateway, as described by the reference architecture, has a double role, as on the one hand it is responsible to relay the measurement data from sensors to edge node, and on the other hand to relay the control data from the edge node to the control cube responsible for the watering process. 

\subsection{Performance evaluation} \label{sec::bigper}

In the first greenhouse (traditional irrigation) the process is supervised by the farmer twice a day for a couple of hours per time, and the control is based on his experience. On the other hand, the second greenhouse is managed by the developed SIN system. Based on the measurements of the sensors, the control cube aims at keeping the ground's moisture between $50\%$ to $55\%$, which is the desired ground's moisture for this cultivation according to the input of the specialized agriculturist of the plantation.

%

The testing period of the SIN system led to some very interesting and efficient results. In Fig.~\ref{fig:measurements} we present the measurements of a 3 days time window which can equally represent the quality outcomes for the rest, extended period samples. The results include the measurements coming from both the traditional irrigation method and the SIN system. In Fig.~\ref{fig:moisture}, we present the results regarding the soil moisture of the greenhouses. In Fig.~\ref{fig:water}, we present the results regarding the water consumption given a fixed water pressure. 

Its is visible that our SIN achieves a soil moisture with much less oscillations than the traditional one. This is due to intense turnovers between drought and high soil moisture levels for the traditional case, in which the farmer enables the watering process. Also, the SIN, contrary to the traditional irrigation method, maintains the soil moisture values within the desired levels. Consequently, the traditional activity deals with an inefficient water usage which finally increases the production costs. On the other hand, in the greenhouse managed by the SIN, the soil moisture remains almost stable for the 3 day period. This is because the watering process is frequently activated to sustain the proper levels of moisture. The hours of watering (hours that the valves remain open with a constant flow of water) required to achieve the aforementioned results are displayed in Fig.~\ref{fig:water}. For the SIN case we achieve more than $50\%$ less hours of operating the watering valves, which also implies an equivalent water saving improvement. Finally, an additional benefit introduced by the SIN usage is the decrease in the number of person months required by the farmer to manage onsite the irrigation process. This is due to the fact that control and monitoring is able to perform the corresponding actuation unsupervised.

\subsection{Cost-benefit considerations}

Based on our deployment experience, further to the performance results, we now provide some insights about the estimation of the strengths and weaknesses of the two alternative SIN management approaches presented in section \ref{sec::big}. Our comments highlight that the smart irrigation approach can achieve greater benefits while preserving significantly higher savings than the manual irrigation approach. The main goal of an irrigation approach is to preserve efficient levels of expenditure while at the same time to provide high production benefits. This goal can be achieved by minimizing the long term costs related to the installation and the continuous operation of the irrigation facility, while maximizing the inherent benefits of each selected irrigation approach. Regarding the costs, on the one hand, the manual irrigation approach requires a greater investment in finding and employing on a regular basis specialized personnel for performing manually the irrigation tasks. Also, as demonstrated by the performance evaluation, the manual approach results in significantly higher water consumption costs, which can quickly escalate in large-scale strawberry plantations. On the other hand, the smart irrigation approach necessitates the purchase and (one-shot) installation of the related IoT equipment described in sections \ref{sec::big}. Also, if the backhaul communication option with the global cloud is considered for additional functionality, we have to take into account the costs derived by the network connectivity and access. In many cases, and especially in remote rural areas with low connectivity and harsh terrains, those costs can significantly increase and require big investments. This setting can be particularly relevant to emerging product tracking mechanisms using, for example, the blockchain paradigm, whereby transaction records grow immensely in size, are not able to be stored locally on the edge of the network, and have to be moved to the global cloud. However, regardless of those costs, the smart irrigation approach comes with significant benefits which are absent from the manual irrigation approach. At first, as demonstrated in section \ref{sec::bigper}, the balancing of ground moisture can be significantly enhanced, leading to a final product of increased quality. Secondly, the smart irrigation solution can be highly scalable, leading to a convenient ease of replication of the irrigation process to numerous greenhouses and facilities. Last but not least, the data collection, which is an outcome of the smart irrigation process, can provide valuable insights as input for additional edge data analytics targeting business decision making; an input that in the manual irrigation case comes just from the personal experience of the farmer.

\section{Conclusions} \label{sec::conc}

In this paper, contrary to the traditional cloud-based solutions, we propose a decentralized smart irrigation approach for strawberry greenhouses which keeps the agricultural data at the edge of the network. After having developed a small-scale smart irrigation networking prototype system and having designed a reference architecture targeting edge data distribution for strawberry greenhouse applications, we implement a full-scale smart irrigation system in an actual strawberry greenhouse environment in Greece. We then compare the performance of our approach to the performance of  conventional irrigation methods managed by the farmer. We conclude that our smart irrigation approach significantly outperforms the conventional approach both in terms of soil moisture variation and in terms of water consumption.

\section*{References}



\begin{thebibliography}{10}
\expandafter\ifx\csname url\endcsname\relax
  \def\url#1{\texttt{#1}}\fi
\expandafter\ifx\csname urlprefix\endcsname\relax\def\urlprefix{URL }\fi
\expandafter\ifx\csname href\endcsname\relax
  \def\href#1#2{#2} \def\path#1{#1}\fi

\bibitem{7879140}
M.~{Roopaei}, P.~{Rad}, K.~R. {Choo}, Cloud of things in smart agriculture:
  Intelligent irrigation monitoring by thermal imaging, IEEE Cloud Computing
  4~(1) (2017) 10--15.
\newblock \href {https://doi.org/10.1109/MCC.2017.5}
  {\path{doi:10.1109/MCC.2017.5}}.

\bibitem{7469991}
W.~{Shi}, S.~{Dustdar}, The promise of edge computing, Computer 49~(5) (2016)
  78--81.
\newblock \href {https://doi.org/10.1109/MC.2016.145}
  {\path{doi:10.1109/MC.2016.145}}.

\bibitem{s18082611}
T.~P. Raptis, A.~Passarella, M.~Conti,
  \href{http://www.mdpi.com/1424-8220/18/8/2611}{Performance analysis of
  latency-aware data management in industrial iot networks}, Sensors 18~(8)
  (2018).
\newblock \href {https://doi.org/10.3390/s18082611}
  {\path{doi:10.3390/s18082611}}.
\newline\urlprefix\url{http://www.mdpi.com/1424-8220/18/8/2611}

\bibitem{BARTLETT2015127}
A.~Bartlett, A.~Andales, M.~Arabi, T.~Bauder,
  \href{http://www.sciencedirect.com/science/article/pii/S0168169914003317}{A
  smartphone app to extend use of a cloud-based irrigation scheduling tool},
  Computers and Electronics in Agriculture 111 (2015) 127 -- 130.
\newblock \href {https://doi.org/https://doi.org/10.1016/j.compag.2014.12.021}
  {\path{doi:https://doi.org/10.1016/j.compag.2014.12.021}}.
\newline\urlprefix\url{http://www.sciencedirect.com/science/article/pii/S0168169914003317}

\bibitem{7389138}
N.~{Sales}, O.~{Remedios}, A.~{Arsenio}, Wireless sensor and actuator system
  for smart irrigation on the cloud, in: 2015 IEEE 2nd World Forum on Internet
  of Things (WF-IoT), 2015, pp. 693--698.
\newblock \href {https://doi.org/10.1109/WF-IoT.2015.7389138}
  {\path{doi:10.1109/WF-IoT.2015.7389138}}.

\bibitem{LOZANO201644}
D.~Lozano, N.~Ruiz, P.~Gavilan,
  \href{http://www.sciencedirect.com/science/article/pii/S037837741630049X}{Consumptive
  water use and irrigation performance of strawberries}, Agricultural Water
  Management 169 (2016) 44 -- 51.
\newblock \href {https://doi.org/https://doi.org/10.1016/j.agwat.2016.02.011}
  {\path{doi:https://doi.org/10.1016/j.agwat.2016.02.011}}.
\newline\urlprefix\url{http://www.sciencedirect.com/science/article/pii/S037837741630049X}

\bibitem{8229715}
R.~{Carrasco}, I.~{Soto}, F.~{Seguel}, L.~{Osorio-Valenzuela}, C.~{Lagos},
  C.~{Flores}, Water balance analysis in plantations of strawberries, in the
  commune of san pedro, in: 2017 CHILEAN Conference on Electrical, Electronics
  Engineering, Information and Communication Technologies (CHILECON), 2017, pp.
  1--5.
\newblock \href {https://doi.org/10.1109/CHILECON.2017.8229715}
  {\path{doi:10.1109/CHILECON.2017.8229715}}.

\bibitem{8058860}
S.~{Daskalakis}, J.~{Kimionis}, A.~{Collado}, M.~M. {Tentzeris},
  A.~{Georgiadis}, Ambient fm backscattering for smart agricultural monitoring,
  in: 2017 IEEE MTT-S International Microwave Symposium (IMS), 2017, pp.
  1339--1341.
\newblock \href {https://doi.org/10.1109/MWSYM.2017.8058860}
  {\path{doi:10.1109/MWSYM.2017.8058860}}.

\bibitem{8423620}
S.~N. {Daskalakis}, G.~{Goussetis}, S.~D. {Assimonis}, M.~M. {Tentzeris},
  A.~{Georgiadis}, A uw backscatter-morse-leaf sensor for low-power
  agricultural wireless sensor networks, IEEE Sensors Journal 18~(19) (2018)
  7889--7898.
\newblock \href {https://doi.org/10.1109/JSEN.2018.2861431}
  {\path{doi:10.1109/JSEN.2018.2861431}}.

\bibitem{4382216}
S.~{Yoo}, J.~{Kim}, T.~{Kim}, S.~{Ahn}, J.~{Sung}, D.~{Kim}, A2s: Automated
  agriculture system based on wsn, in: 2007 IEEE International Symposium on
  Consumer Electronics, 2007, pp. 1--5.
\newblock \href {https://doi.org/10.1109/ISCE.2007.4382216}
  {\path{doi:10.1109/ISCE.2007.4382216}}.

\bibitem{4340413}
Y.~{Zhou}, X.~{Yang}, X.~{Guo}, M.~{Zhou}, L.~{Wang}, A design of greenhouse
  monitoring control system based on zigbee wireless sensor network, in: 2007
  International Conference on Wireless Communications, Networking and Mobile
  Computing, 2007, pp. 2563--2567.
\newblock \href {https://doi.org/10.1109/WICOM.2007.638}
  {\path{doi:10.1109/WICOM.2007.638}}.

\bibitem{inproceedings}
M.~Hebel, Meeting wide-area agricultural data acquisition and control
  challenges through zigbee wireless network technology, in: Computers in
  Agriculture and Natural Resources, 4th World Congress Conference, 2006.
\newblock \href {https://doi.org/10.13031/2013.21879}
  {\path{doi:10.13031/2013.21879}}.

\bibitem{4457920}
Y.~{Kim}, R.~G. {Evans}, W.~M. {Iversen}, Remote sensing and control of an
  irrigation system using a distributed wireless sensor network, IEEE
  Transactions on Instrumentation and Measurement 57~(7) (2008) 1379--1387.
\newblock \href {https://doi.org/10.1109/TIM.2008.917198}
  {\path{doi:10.1109/TIM.2008.917198}}.

\bibitem{7133593}
M.~{Rivers}, N.~{Coles}, H.~{Zia}, N.~R. {Harris}, R.~{Yates}, How could sensor
  networks help with agricultural water management issues? optimizing
  irrigation scheduling through networked soil-moisture sensors, in: 2015 IEEE
  Sensors Applications Symposium (SAS), 2015, pp. 1--6.
\newblock \href {https://doi.org/10.1109/SAS.2015.7133593}
  {\path{doi:10.1109/SAS.2015.7133593}}.

\bibitem{7133592}
H.~{Zia}, N.~{Harris}, G.~{Merrett}, M.~{Rivers}, Data-driven low-complexity
  nitrate loss model utilizing sensor information - towards collaborative
  farm management with wireless sensor networks, in: 2015 IEEE Sensors
  Applications Symposium (SAS), 2015, pp. 1--6.
\newblock \href {https://doi.org/10.1109/SAS.2015.7133592}
  {\path{doi:10.1109/SAS.2015.7133592}}.

\bibitem{7930391}
F.~{Viani}, M.~{Bertolli}, M.~{Salucci}, A.~{Polo}, Low-cost wireless
  monitoring and decision support for water saving in agriculture, IEEE Sensors
  Journal 17~(13) (2017) 4299--4309.
\newblock \href {https://doi.org/10.1109/JSEN.2017.2705043}
  {\path{doi:10.1109/JSEN.2017.2705043}}.

\bibitem{8612441}
P.~{Serikul}, N.~{Nakpong}, N.~{Nakjuatong}, Smart farm monitoring via the
  blynk iot platform : Case study: Humidity monitoring and data recording, in:
  2018 16th International Conference on ICT and Knowledge Engineering (ICT KE),
  2018, pp. 1--6.
\newblock \href {https://doi.org/10.1109/ICTKE.2018.8612441}
  {\path{doi:10.1109/ICTKE.2018.8612441}}.

\bibitem{8644215}
M.~M. {Maha}, S.~{Bhuiyan}, M.~{Masuduzzaman}, Smart board for precision
  farming using wireless sensor network, in: 2019 International Conference on
  Robotics,Electrical and Signal Processing Techniques (ICREST), 2019, pp.
  445--450.
\newblock \href {https://doi.org/10.1109/ICREST.2019.8644215}
  {\path{doi:10.1109/ICREST.2019.8644215}}.

\bibitem{7219582}
S.~{Jindarat}, P.~{Wuttidittachotti}, Smart farm monitoring using raspberry pi
  and arduino, in: 2015 International Conference on Computer, Communications,
  and Control Technology (I4CT), 2015, pp. 284--288.
\newblock \href {https://doi.org/10.1109/I4CT.2015.7219582}
  {\path{doi:10.1109/I4CT.2015.7219582}}.

\bibitem{8644296}
A.~{Mukherjee}, N.~{Pathak}, S.~{Misra}, S.~{Mitra}, Predictive intra-edge
  packet-source mapping in agricultural internet of things, in: 2018 IEEE
  Globecom Workshops (GC Wkshps), 2018, pp. 1--6.
\newblock \href {https://doi.org/10.1109/GLOCOMW.2018.8644296}
  {\path{doi:10.1109/GLOCOMW.2018.8644296}}.

\bibitem{BU2019500}
F.~Bu, X.~Wang,
  \href{http://www.sciencedirect.com/science/article/pii/S0167739X19307277}{A
  smart agriculture iot system based on deep reinforcement learning}, Future
  Generation Computer Systems 99 (2019) 500 -- 507.
\newblock \href {https://doi.org/https://doi.org/10.1016/j.future.2019.04.041}
  {\path{doi:https://doi.org/10.1016/j.future.2019.04.041}}.
\newline\urlprefix\url{http://www.sciencedirect.com/science/article/pii/S0167739X19307277}

\bibitem{Chi2019}
T.~Chi, M.~Chen, \href{https://doi.org/10.1007/s11276-017-1593-z}{A frequency
  hopping method for spatial rfid/wifi/bluetooth scheduling in agricultural
  iot}, Wireless Networks 25~(2) (2019) 805--817.
\newblock \href {https://doi.org/10.1007/s11276-017-1593-z}
  {\path{doi:10.1007/s11276-017-1593-z}}.
\newline\urlprefix\url{https://doi.org/10.1007/s11276-017-1593-z}

\bibitem{Angelopoulos2011}
C.~M. Angelopoulos, S.~Nikoletseas, G.~C. Theofanopoulos,
  \href{http://doi.acm.org/10.1145/2069131.2069162}{A smart system for garden
  watering using wireless sensor networks}, in: Proceedings of the 9th ACM
  International Symposium on Mobility Management and Wireless Access, MobiWac
  '11, ACM, New York, NY, USA, 2011, pp. 167--170.
\newblock \href {https://doi.org/10.1145/2069131.2069162}
  {\path{doi:10.1145/2069131.2069162}}.
\newline\urlprefix\url{http://doi.acm.org/10.1145/2069131.2069162}

\bibitem{8764545}
T.~P. {Raptis}, A.~{Passarella}, M.~{Conti}, Data management in industry 4.0:
  State of the art and open challenges, IEEE Access 7 (2019) 97052--97093.
\newblock \href {https://doi.org/10.1109/ACCESS.2019.2929296}
  {\path{doi:10.1109/ACCESS.2019.2929296}}.

\bibitem{Lucas2018}
M.~Lucas-Estan, M.~Sepulcre, T.~Raptis, A.~Passarella, M.~Conti,
  \href{http://dx.doi.org/10.3390/electronics7120400}{Emerging trends in hybrid
  wireless communication and data management for the industry 4.0}, Electronics
  7~(12) (2018) 400.
\newblock \href {https://doi.org/10.3390/electronics7120400}
  {\path{doi:10.3390/electronics7120400}}.
\newline\urlprefix\url{http://dx.doi.org/10.3390/electronics7120400}

\bibitem{Hosseini2019}
M.~Hosseini, C.~M. Angelopoulos, W.~K. Chai, S.~Kundig,
  \href{https://doi.org/10.1007/s10586-018-2843-2}{Crowdcloud: a crowdsourced
  system for cloud infrastructure}, Cluster Computing 22~(2) (2019) 455--470.
\newblock \href {https://doi.org/10.1007/s10586-018-2843-2}
  {\path{doi:10.1007/s10586-018-2843-2}}.
\newline\urlprefix\url{https://doi.org/10.1007/s10586-018-2843-2}

\bibitem{6655569}
C.~M. {Angelopoulos}, G.~{Filios}, S.~{Nikoletseas}, D.~{Patroumpa}, T.~P.
  {Raptis}, K.~{Veroutis}, A holistic ipv6 test-bed for smart, green buildings,
  in: 2013 IEEE International Conference on Communications (ICC), 2013, pp.
  6050--6054.
\newblock \href {https://doi.org/10.1109/ICC.2013.6655569}
  {\path{doi:10.1109/ICC.2013.6655569}}.

\bibitem{6846177}
S.~{Nikoletseas}, M.~{Rapti}, T.~P. {Raptis}, K.~{Veroutis}, Decentralizing and
  adding portability to an iot test-bed through smartphones, in: 2014 IEEE
  International Conference on Distributed Computing in Sensor Systems, 2014,
  pp. 281--286.
\newblock \href {https://doi.org/10.1109/DCOSS.2014.39}
  {\path{doi:10.1109/DCOSS.2014.39}}.

\bibitem{ec-5}
Decagon ec-5:
  \url{http://www.decagon.com/products/sensors/soil-moisture-sensors/ec-5-soil-moisture-small-area-of-influence/}.

\end{thebibliography}

\end{document}